\documentclass[aps,prd,twocolumn,preprintnumbers,nofootinbib,showpacs]{revtex4-1}
\usepackage{graphicx}
\usepackage{amssymb}
\usepackage{amsmath}
\usepackage{color}
\usepackage{enumerate}
\usepackage{overpic}

\def\beq{\begin{equation}}
\def\eeq{\end{equation}}
\def\beqn{\begin{eqnarray}}
\def\eeqn{\end{eqnarray}}

\renewcommand{\texttt}{{}}
\newcommand{\be}{\begin{eqnarray}}
\newcommand{\ee}{\end{eqnarray}}
\newcommand{\bee}{\begin{equation}}
\newcommand{\eee}{\end{equation}}

\begin{document}

\title{Singularity avoidance in classical gravity from four-fermion interaction}

\author{Cosimo Bambi}
\affiliation{Center for Field Theory and Particle Physics \& Department of Physics, 
Fudan University, 200433 Shanghai, China}

\author{Daniele Malafarina}
\affiliation{Center for Field Theory and Particle Physics \& Department of Physics, 
Fudan University, 200433 Shanghai, China}

\author{Antonino Marcian\`o}
\affiliation{Center for Field Theory and Particle Physics \& Department of Physics, 
Fudan University, 200433 Shanghai, China}

\author{Leonardo Modesto}
\affiliation{Center for Field Theory and Particle Physics \& Department of Physics, 
Fudan University, 200433 Shanghai, China}

\date{\today}

\begin{abstract}
\noindent 
We derive the dynamics of the gravitational collapse of a homogeneous and spherically symmetric cloud in a classical set-up endowed with a topological sector of gravity and a non-minimal coupling to fermions. The effective theory consists of the Einstein-Hilbert action plus Dirac fermions interacting through a four-fermion vertex. At the classical level, we obtain the same picture that has been recently studied by some of us within a wide range of effective theories inspired by a super-renormalizable and asymptotically free theory of gravity. The classical singularity is replaced by a bounce, beyond which the cloud re-expands indefinitely. We thus show that, even at a classical level, if we allow for a non-minimal coupling of gravity to fermions, event horizons may never form for a suitable choice of some parameters of the theory. 
\end{abstract}


\maketitle

\noindent
In a previous work, some of us have studied the gravitational collapse in a wide class of asymptotically free theories of gravity~\cite{p}. It was found a picture that substantially differs from the standard scenario. The central singularity that appears in classical general relativity is replaced by a bounce, after which the collapsing body starts expanding. It was argued that, strictly speaking, black holes never form, in the sense that there are no regions causally disconnected to future null infinity.
The collapse can only produce a temporary trapped surface, which looks like an event horizon for an observational timescale much shorter than the one of the collapse. While this time interval is of order a dynamical timescale for a comoving observer, it is definitively long for an observer in the exterior metric that is far away from the collapsing body. For all practical purposes these objects are therefore like black holes. Similar studies have been presented in Ref.~\cite{pc}. In the present work, we show that the same picture can be found in classical general relativity, when we extend the gravitational sector to include topological terms and we consider an experimentally allowed non-minimal coupling of fermions in the Dirac action.

We can start from the non-minimal Einstein-Cartan-Holst (ECH) action, as cast by Bojowald and Das in \cite{BD}
\begin{eqnarray}
\label{nonminimalaction}
\!\!\!\!&& S \left[e,A,\psi\right] = S_{G}\left[e,A \right]+
S_{F}\left[   e, A ,\psi  \right] = \\
\!\!\!\!&& = \frac{1}{2 \kappa} \int d^{4}x \;|e|e^{\mu}_{I}e^{\nu}_{J}
P^{IJ}_{\ \ \ KL}F^{\ \ KL}_{\mu
\nu}(A) + \nonumber\\
\!\!\!\!&& + \frac{i}{2}  \int \!\! d^{4}x 
|e| \left[\overline{\psi}\gamma^{I}e^{\mu}_{I}\left(1-
\frac{i }{\alpha}\gamma_{5}\right)\nabla_{\mu}\psi 
+ i  \, m \overline{\psi} \psi + {\rm h.c.}\right]   ,
\nonumber
\end{eqnarray}
where $\kappa = 8 \pi G_{\rm N}$ is the reduced Planck length square. Notice the presence of a non-minimal coupling parameter $\alpha\in \mathbb{R}$, which has been first introduced by Freidel, Minic and Takeuchi in \cite{Freidel:2005sn}, but without $\gamma_5$. This $\gamma_5$ turns out to be crucial for parity invariance and was introduced by Mercury in~\cite{mercuri}. The experimental bounds for $\alpha$ and $\gamma$ arising from lepton-quark contact interactions are discussed in~\cite{Freidel:2005sn}. 
The operator
\begin{equation}
\label{PIJ}
P^{IJ}_{\ \ \ KL}=\delta^{[I}_{K} \delta^{J]}_{L} - \frac{1}{2 \gamma} \epsilon^{IJ}_{\ \ KL} \, ,
\end{equation}
where $\epsilon_{IJKL}$ is the Levi-Civita symbol, 
is defined in terms of the Barbero--Immirzi parameter $\gamma$, and can be inverted for $\gamma^2\neq - 1$. As shown in~\cite{mercuri}, the Einstein-Cartan action is recovered for $\alpha=\gamma$, with a term that reduces to the Nieh-Yan invariant when the second Cartan structure equation holds. This case is referred to as minimal coupling in the Einstein-Cartan theory. From the point of view of the Holst action, minimal coupling is met in the limit $\alpha \rightarrow \pm\infty$.

The covariant derivative $\nabla_{\mu}$ of Dirac spinors and the field-strength of the Lorentz connection are defined by
\begin{equation} \label{covariantderivative}
 \nabla_{\mu}\equiv \partial_{\mu} + \frac{1}{4}A^{IJ}_{\mu}
 \gamma_{[I} \gamma_{J]}\, , \quad
 \left[\nabla_{\mu},\nabla_{\nu}\right] = \frac{1}{4}F^{IJ}_{\mu
 \nu}\gamma_{[I} \gamma_{J]}\,.
\end{equation}

Because of the presence of fermions, a torsional part of the connection enters the non-minimal ECH action. Nevertheless, we can follow here the procedure used by Perez and Rovelli in \cite{PR}, and integrate out of the theory the torsional part of the connection through the Cartan equation, which is found by varying the total action with respect to the connection $A^{IJ}_\mu$.
The variation of the action with respect to the connection $A$ gives
\be
P^{I J}\,_{K L} \nabla_\mu ( e e^\mu_I e^\nu_J ) = \kappa \, e \,J_{K L}^\nu \,.
\label{eqA}
\ee
This equation can then be solved for the connection. For this purpose, we write the connection in the form 
\be
A_\mu^{IJ} = \omega(e)^{I J}_\mu + C_\mu^{I J}, 
\label{A}
\ee
where $C_\mu^{I J}$ is the contorsion tensor and $\omega(e)$ is the torsion free spin connection determined by $e$, namely the solution of $\widetilde{\nabla}_{[\mu}  e_{\nu]}^I=0$. Note that we have introduced a new definition for the covariant derivative compatible with the tetrad $e^I_\mu$,
\be
\widetilde{\nabla}_{\mu}\equiv \partial_{\mu} + \frac{1}{4}\omega^{IJ}_{\mu} \gamma_{[I} \gamma_{J]}\, .
\ee
Replacing the definition (\ref{A}) in (\ref{eqA}) we find
\be
C_{\mu [I}^\mu e_{J]}^\nu + C_{[I J]}^\nu = \kappa \, (P^{-1})_{I J}\,^{K L}e \,J_{K L}^\nu \, , \label{eqA2}
\ee
in which
\be
\hspace{-0.4cm}
J_{KL}^\nu= e\,\frac{1}{4} e^\nu_I \epsilon^I_{\ KLM} \overline{\psi} \gamma_5 \gamma^M \psi - \frac{1}{2 \alpha} e^{\nu I}\, \eta_{I [K} \, \overline{\psi} \gamma_5 \gamma_{L]} \psi \,.
\ee
Note that we transform internal and spacetime indices into one another, using the tetrad field, and preserving the horizontal order of the indices. 
Then the Cartan equation expresses the contortion tensor $C_\mu^{IJ}$ in terms of the fermionic fields and tetrads
\be \label{carta}
e^\mu_I \, C_{\mu JK} = \frac{\kappa}{4} \, \frac{\gamma}{\gamma^2 +1} \, \left( \beta \, \epsilon _{IJKL}\ J^L - 2 \theta \, \eta_{I[J} \, J_{K]}  \right)  , 
\ee
having introduced the flat metric $\eta_{IJ}$, the fermionic axial current $J^L=\overline{\psi} \gamma^L \gamma_5 \psi$, and the coefficients, functions of the free parameters within the non-minimal ECH theory, $\beta=\gamma+1/\alpha$ and $\theta=1-\gamma/\alpha$. Thanks to (\ref{carta}) the non-minimal ECH action recasts in terms of the metric compatible connection, as a sum of the Einstein-Hilbert action and the Dirac action. The latter is now written in terms of metric compatible variables, and is further provided with novel interaction terms, which capture the new physics within the non-minimal ECH theory. Consequences of this new interaction term in cosmology have been investigated by Alexander, Biswas and Calcagni in \cite{ABC}. The theory can be rewritten as
\begin{eqnarray}
\label{nonminimalaction2}
&&S \left[e,A,\psi\right] = S_{G}\left[e,\omega \right]+
S_{F}\left[e, \omega ,\psi  \right]  + S_{\rm int}[e, \psi] = \nonumber\\
&&= \frac{1}{2 \kappa} \int d^{4}x |e|e^{\mu}_{I}e^{\nu}_{J}
F^{I J }_{\mu\nu}(\omega) + \nonumber\\
&&+ \frac{i}{2}  \int d^{4}x \, |e| \Big(\overline{\psi}\gamma^{I}e^{\mu}_{I}
\nabla_{\mu}\psi - \overline{\nabla_{\mu} \psi}\gamma^{I}e^{\mu}_{I}
\psi + i  \, m \overline{\psi} \psi  \Big) + \nonumber \\
&& - \kappa \, \xi \int d^{4}x \, | e | (\bar{\psi} \gamma_5 \gamma^L \psi )  
(\bar{\psi} \gamma_5 \gamma_L \psi )  \, ,
\end{eqnarray}
where
\be
\xi = \frac{3}{16} \frac{\gamma^2}{\gamma^2+1} 
\left(1 + \frac{2}{\alpha \gamma} - \frac{1}{\alpha^2}\right) \, .
\ee

Einstein equations $G_{\mu\nu} = \kappa \, T_{\mu\nu}$ provide the dynamics for the gravitational field $e^I_\mu$, and must be coupled to the equations of motion for fermionic matter and radiation. We have denoted with $G_{\mu\nu}$ the Einstein tensor and the stress-energy tensor is 
\be
T_{\mu\nu} \!=\! \frac{e_{\mu\,I} } {|e|} \frac{\delta    \left( |e| \mathcal{L}_{\rm matt} \right)}{\delta e^\nu_I}.
\ee
The fermionic Lagrangian including the interaction reads 
\be
\mathcal{L}_{\rm fer} = | e | \left[ \frac{1}{2} \left( \overline{\psi} \gamma^I e^\mu_I i \widetilde{\nabla}_\mu \psi - m \overline{\psi} \psi \right) \!+ {\rm h.c.} - \kappa  \xi \, J^L J_L \right] \,,
\nonumber
\ee
which yields the energy-momentum tensor 
\be 
&& T^{\rm fer}_{\mu\nu} =  \frac{1}{4} \left(  \overline{\psi} \gamma_I e^I_\mu i  \widetilde{\nabla}_\nu \psi +  \overline{\psi} \gamma_I e_\nu^I i \widetilde{\nabla}_\mu \psi \right) + {\rm h.c.} \nonumber \\
&& \hspace{1cm} - g_{\mu\nu}\, \mathcal{L}_{\rm fer} \,  .  \label{tenfe}
\ee
The Dirac equations on curved background for the interacting system are the following, 
\be
&&\!\!\!\!\!\!\!\!\!\!\!\!\!\!\!\!\!\!\!\!\!\! \gamma^I e_I^\mu i  \widetilde{\nabla}_{\mu} \psi  - m \psi  =\nonumber \\
&&\ \ \ \ =  2 \xi \kappa  (\overline{\psi} \psi +  \overline{\psi} \gamma_5 \psi   \gamma_5
+  \overline{\psi} \gamma_I \psi  \gamma^I) \psi \, ,
\ee     
in which we have used the Fierz-decomposition 
\be
&& \hspace{0cm}
\!\!\!\!\!\!\!\!\!\!\!\!\!\!\!\!\!\!\!\!
(\overline{\psi} \gamma_5 \gamma^I  \psi)  (\overline{\psi} \gamma_5 \gamma_I  \psi)= \nonumber \\
&& = (\overline{\psi}  \psi)^2 
+ (\overline{\psi} \gamma_5  \psi)^2 + (\overline{\psi} \gamma^I \psi)  (\overline{\psi}  \gamma_I  \psi).
\ee

In what follows, we study the dynamics of the collapse of a homogeneous and 
spherically symmetric body. In the comoving gauge, the tetrad $e^I_\mu$ for the 
Friedmann-Lema\^itre-Robertson-Walker (FLRW) type metrics is 
\be 
e^I_0=\delta^I_0 \,\,\, {\mbox{and}} \,\,\, e^I_j= a(t) \, \delta^I_j \, , 
\ee
where $a$ is the FLRW scale factor and $t$ is the comoving time. Solutions of the Dirac equations on curved backgrounds that are suitable to develop cosmological analyses have been studied by Armendariz-Picon and Greene~\cite{AP}. 
They resorted to a form of the spinor which allows for the vanishing of the spatial components of the vector (but not of the axial) fermionic current 
\be
\psi= (\psi_0(t) , 0,0,0) \, .
\ee
This ensures homogeneity and isotropy on spatial hyper-surfaces for theories in which a cooling between vector current and any other observable vector quantity is present. We then simplify the Dirac equation using their {\it ansatz}, which still holds in our framework due to the appearance of only quadratic powers of $J_L$. Within the comoving gauge, the only non-vanishing spin connection components for $\omega^{IJ}_{\;\;\;\, K}\!=\!\omega^{IJ}_\mu\, e^\mu_K$ are $\omega_{0ij}=-\omega_{i0j}= - H \delta_{ij}$. This implies $\widetilde{\nabla}_0 = \partial_0$ and $\widetilde{\nabla}_i = \partial_i+ a H/2 \delta_{ij}{\rm diag} (\sigma^j, - \sigma^j)$. The Dirac equation then reads
\be
\dot{\psi}_0 + \frac{3}{2} \,H\, \psi_0 + i  
\left(m + 4 \, \kappa \, \xi \,  \psi_0^{*} \psi_0 \right) \psi_0=\!0 \, ,
\ee
where $^{*}$ denotes complex conjugation. The equation of motion for the 
bilinear $\psi_0^{*}\psi_0$ is
\be \label{seque}
\frac{d }{dt} \, \psi_0^{*} \psi_0+ 3\,H\, \psi_0^{*} \psi_0 =0\,,
\ee
and yields the familiar $a^{-3}$ scaling for the particle number density
\be \label{ro}
\psi_0^{*} \psi_0 = n_0/a^3\,,
\ee
where $n_0$ is a constant. With the use of Eq.~(\ref{ro}), the first Friedmann equation reads
\be\label{H2}
H^2 = \frac{\kappa \, m}{3}\frac{n_0}{a^3}  
+ \frac{\kappa^2 \, \xi}{3}\,\frac{n_0^2}{a^6} \,.
\ee 
Here the first term on the right hand side is the standard term describing dust matter, while the second one comes from the four-fermions interaction and originates from integrating out the torsionful part of the gravitational connection.

\begin{figure*}[t]
\hspace{0.7cm}
\begin{center}
\begin{overpic}[scale=0.6]{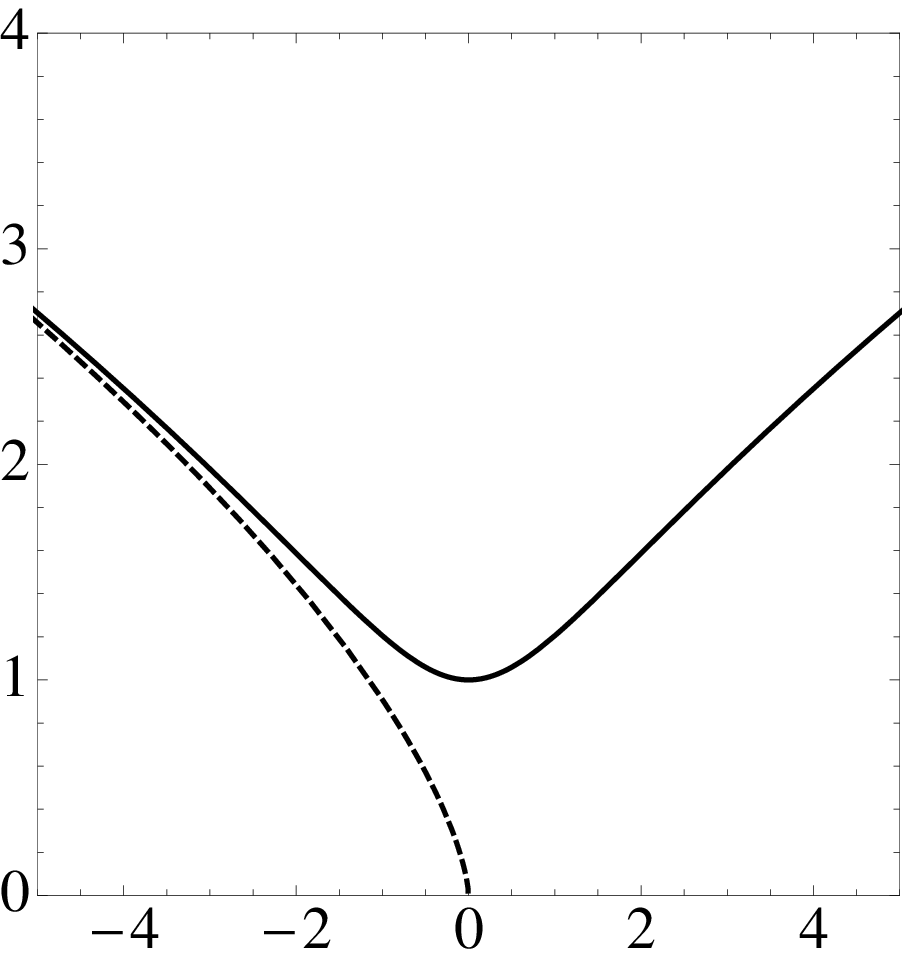}
\put(-10,50){$a$}
\put(50,-6){$t$}
\end{overpic}
\hspace{1cm}
\begin{overpic}[scale=0.625]{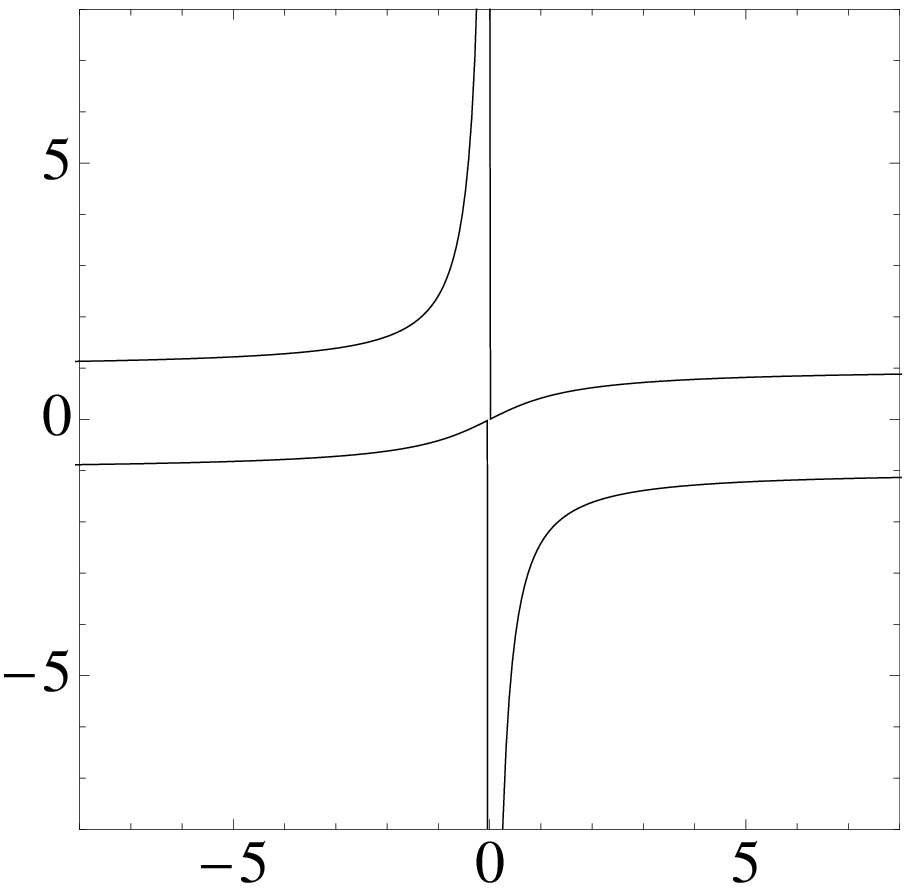}
\put(-6,50){$\alpha$}
\put(50,-6){$\gamma$}
\put(25,75){$\xi>0$}
\put(25,53){$\xi<0$}
\put(25,25){$\xi>0$}
\put(70,75){$\xi>0$}
\put(70,50){$\xi<0$}
\put(70,25){$\xi>0$}
\end{overpic}
\end{center}
\vspace{0.5cm}
\begin{center}
\hspace{-0.8cm}
\begin{overpic}[scale=0.67]{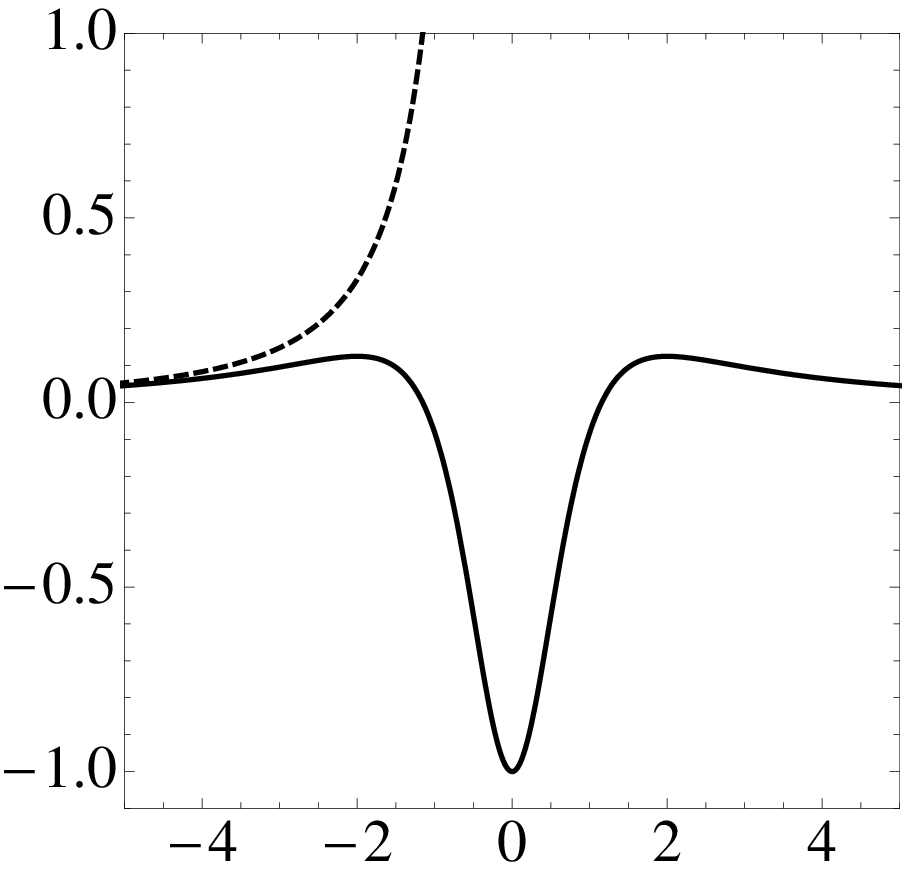}
\put(-15,60){$\rho_{\rm eff} + p_{\rm eff}$}
\put(55,-6){$t$}
\end{overpic}
\hspace{1.1cm}
\begin{overpic}[scale=0.605]{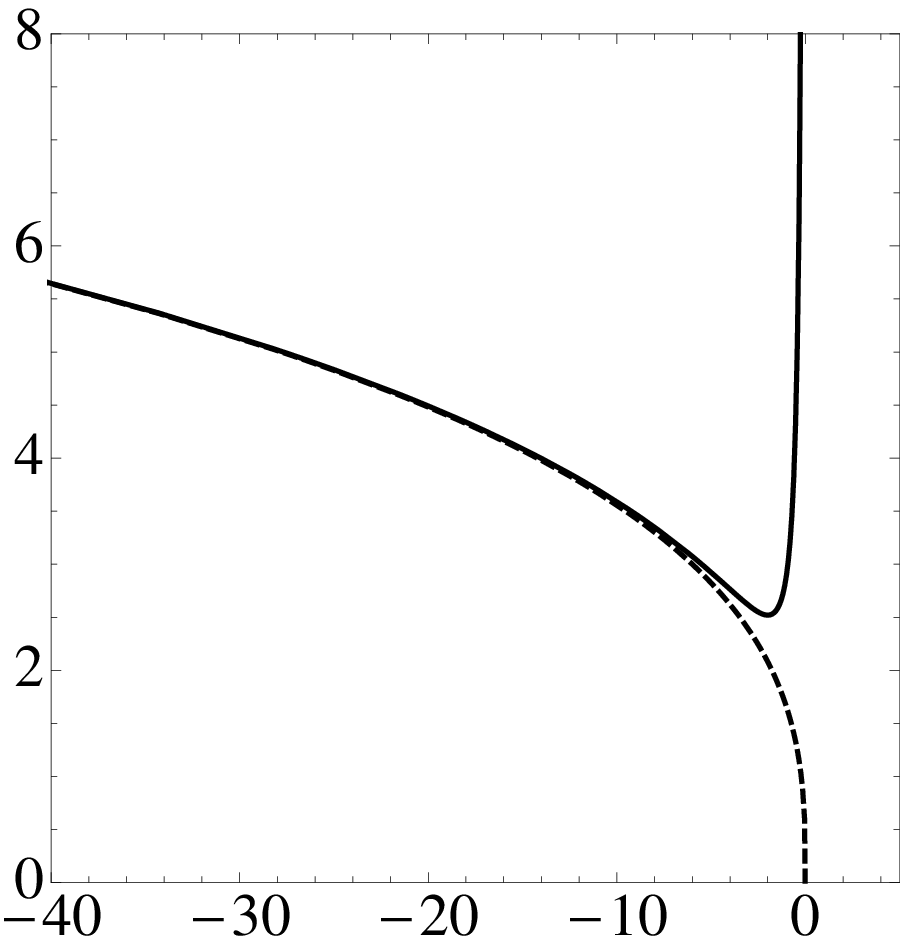}
\put(-10,50){$r_{\rm ah}$}
\put(50,-6){$t$}
\end{overpic}
\end{center}
\caption{Top left panel: scale factor $a(t)$ for the non-minimal coupling collapse 
(solid line) and the standard scenario (dashed line). Top right panel: sign of 
$\xi$ on the plane ($\gamma$,$\alpha$). 
Bottom left panel: $\rho_{\rm eff} + p_{\rm eff}$ (solid line) and 
energy density $\rho$ for the classical dust case (dashed line).
Bottom right panel: evolution of the apparent 
horizon $r_{\rm ah}(t)$ for the non-minimal coupling collapse (solid line) 
and the standard scenario (dashed line). In both the left and the right panels, 
$\kappa = m= n_0=\xi =1$. See the text for more details.} 
\label{fig} 
\end{figure*}

The dynamics of the system is inevitably governed by the standard term at lower densities, and by the new one at higher densities. The crucial point is that the sign of $\xi$ depends on the values of $\alpha$ and $\gamma$, and when it is negative we have a bounce. At the time of the bounce $t=t_{\rm B}$, $H$ vanishes, the scale factor reaches its minimum $a_{\rm B}=(- \kappa \xi n_0/m)^{1/3}$, while the bilinear $\psi_0^{*}\psi_0$ reaches its maximum. The scale factor is recovered to be
\be
a(t)=\left[ \frac{3\, m \, \kappa\, n_0}{4} (t-t_{\rm B})^2 - 
 \frac{\kappa \, \xi \, n_0}{m}  \right]^{1/3} \,.
\ee
This solution is shown to be stable under perturbations to the fermionic matter field, if  the anisotropic and inhomogeneous contribution to the energy density, which reads
\be
\tilde{\rho} \sim \frac{ \rm{Tr}{[\gamma_i \gamma_j]} }{M_p^2} \, \overline{\psi} \psi \, \langle \delta \overline{\psi} \delta \psi \rangle
\,,
\ee
is subdominant with respect to the isotropic contribution in the second hand side of (\ref{H2}). The criterion to have a subdominant contribution casts as
\be
 \langle \delta \overline{\psi} \delta \psi \rangle/M_p^2 <\!\!< m\,. 
\ee
The contribution from $\tilde{\rho}$ to the Friedmann equations can be evaluated using the solutions to the in-homogenous and anisotropic perturbations to the fermionic field, as recovered in \cite{ABMM}, which finally provides an explicit condition as developed in \cite{ACM}, namely
\be
m >\!\!> 4 \, n_0/M_p^2\,.
\ee 
The scale factor $a(t)$ is shown in the top left panel of Fig.~\ref{fig} and it is compared with the scale factor of the standard and singular scenario $\kappa\,\xi \rightarrow 0$. The sign of $\xi$ on the plane ($\gamma$,$\alpha$) is shown in the top right panel of Fig.~\ref{fig}. The presence of the bounce and the resolution of the classical singularity is only determined by the sign of $\xi$, not by its numerical value.

The dynamics of the collapse can then be described by an effective model, whose evolutionary equations are derived from Einstein's gravity coupled to a perfect fluid. It is convenient to cast the effective energy density $\rho_{\rm eff}$ as the sum of the physical energy density (in our case, the energy density of dust) plus a correction, which must be small at low densities and becomes relevant when new physics appears. In our case
$\rho_{\rm eff}=\rho_{\rm dust}+\rho_{\rm corr}$, where
\be
\rho_{\rm dust} = m \, \frac{n_0}{a^3} \, , \quad {\rm and}
\quad \rho_{\rm corr} =  \kappa \,\xi\,  \frac{n_0^2}{a^6} \, .
\ee
We write $\rho_{\rm eff}$ as customary in quantum cosmology as~\cite{book}
\be\label{eff}
\rho_{\rm eff}=\rho_{\rm dust}\left(1-\frac{\rho_{\rm dust}}{\rho_{\rm cr}}\right)\,,
\ee
in which $\rho_{\rm cr}=- m^2/(\kappa\, \xi)$ is the critical density at which the non-minimal fermionic coupling becomes relevant. 
The qualitative picture for the gravitational collapse is the same as the one studied in~\cite{p}, but since here we are dealing with a classical effect, although arising from the quantum nature of the fields, we expect the critical density to be much lower than the Planck density at which quantum-gravity effects are supposed to show up. Therefore we obtain a classical bounce that occurs before reaching the energy scales proper to quantum-gravity phenomena.

The origin of the bounce and the avoidance of the classical singularity can be easily understood in this effective picture from the second Friedmann equation:
\be\label{Hpunto}
\hspace{-0.5cm}
\frac{\ddot{a}}{a}=  -  \frac{\kappa}{6} \left(\rho_{\rm eff} + 3 p_{\rm eff}\right)
= -  \frac{\kappa}{6} \left(\! m \frac{n_0}{a^3} + 4 \, \kappa \,\xi \, \frac{n_0^2}{a^6}  \right)\, .
\ee 
While the effective energy density goes to zero as the bounce is approached,
and it is exactly zero at the bounce, we have a negative ($\xi < 0$) 
effective pressure 
\be
p_{\rm eff} = \kappa \, \xi \, \frac{n_0^2}{a^6} \,, 
\ee
which reaches its maximum at the bounce. It is this effective negative pressure that is responsible for the bounce. The formation of the central singularity can indeed be avoid because the weak energy condition $\rho_{\rm eff} + p_{\rm eff} \ge 0$ is violated at a certain point of the evolution of the collapse, see the bottom left panel in Fig.~\ref{fig}.

Unlike previous studies in the literature~\cite{p, pc}, here the bounce arises within the framework of classical gravity and therefore the energy scale is not regulated by the Planck mass. This opens a chance to have some observational implications in high energy astrophysical phenomena. For instance, there might exist objects that are smaller, denser and less massive than neutron stars and are not black holes. Another possibility is that there might exist a threshold that prevents gravitational collapse to take place below the neutron degeneracy pressure but above the black hole formation threshold, which can be  relevant for supernova explosions. Thus such a mechanism might result in sourcing extremely energetic explosive phenomena. Relying on the parameters $\gamma$ and $\alpha$ within the theory, we may try to estimate scales and sizes of the relevant physical processes involved during collapse, with the purpose of figuring out whether the signatures of the non-minimal coupling (regulated by $\alpha$) and of the topological gravitational term (governed by $\gamma$) can eventually be captured from astrophysical observations. This may provide marginal constraints on $\xi$, and therefore on a proper combination of $\gamma$ and $\alpha$. 

In order to study how the non-minimal coupling of fermions and the topological term for gravity can affect the formation of astrophysical black holes, we need to analyze the formation of the trapped surface in the collapsing interior. The apparent horizon is a null surface that determines the boundary between the particles that have light-cones confined in the region causally disconnected from the rest of the universe, and the particles that can propagate to far away observers. In the vacuum exterior, the apparent horizon coincides with the event horizon, while inside the cloud the apparent horizon is recovered thanks to the expression derived in~\cite{p}, {\it i.e.}
\begin{eqnarray}
\label{trapped-surfaces}
r_{\rm ah}(t) = \frac{1}{|\dot{a}|} \, .
\end{eqnarray}
Equation (\ref{trapped-surfaces}) describes the time at which the radius $r$ becomes trapped. In the right panel of Fig.~ \ref{fig}, we show the evolution of the radius of the apparent horizon, $r_{\rm ah}$, for the scenario described above, and we compare it with the one of the classical dust scenario. In the standard case there is no way to avoid the creation of the horizon, which forms at the boundary of the collapsing cloud at a time $t < t_{\rm S}$, and reaches $r=0$ at the time of formation of the singularity $t_{\rm S}$. Thus standard homogeneous relativistic dynamics inevitably induces the formation of a black hole, where the central singularity is always hidden behind a horizon and not visible from faraway observers.

The scenario with a non-minimal coupling and a topological term is qualitatively different. Close to the initial time, the evolution mimics that one of the relativistic collapse without corrections. Nevertheless, as the density increases, the coupling of gravity with fermions becomes relevant. This is described in the effective framework by the occurrence of a negative pressure. The contracting cloud reaches a minimum radius, after which matter bounces, turning the collapse into an expansion. The curve $r_{\rm ah}(t)$ follows the classical model in the early stages, but, as the collapse progresses, it reaches a minimum at a time $t_*$ antecedent the time of the bounce. At $t_{*}$ we see that $\dot{a}$ reaches its maximum value $\dot{a}(t_{*})=\dot{a}_{*}$. 

The main consequence of a minimal radius for the apparent horizon is that there exists a limiting boundary radius $r_{*}=1/|\dot{a}_{*}|$, below which no trapped surface forms at any time. Objects provided with boundary radius $r_b\leq r_{*}$ collapse and bounce without forming any trapped surface. Correspondingly, the limiting radius is related to a threshold mass, below which no black holes can form. The values of $M_{*}$ and $r_{*}$ depend on the parameters appearing in the theory, namely $m$, $n_0$, $\gamma$, and $\alpha$, and thus they are expected to be well above the Planck scale. In fact, since the effective description of the collapse given by equation (\ref{eff}) is the same as the one studied in Ref.~\cite{p}, we can quickly evaluate these limiting quantities and find
\be
r_{*} = \left(\frac{16 \sqrt{-\xi}}{\kappa\, m^2 n_0}\right)^{1/3}\!\!, 
\qquad M_{*} = \frac{8 \sqrt{-\xi}}{3 \kappa m}\,.
\ee
Thus for values of $\xi$ large enough one can in principle have that a stellar object of mass $M_*$ will never form any horizon while collapsing. If we assume $\xi \sim 1$ as a natural value and we consider $m$ equal to the nucleon mass, $M_{*} \sim 10^{14}$~g, which is too small to have implications in the contemporary Universe. Nevertheless, there might be implications in the very early Universe and in the production of the so-called primordial black holes.

{\it Summary and conclusion ---}
It is widely believed that space-time singularities that are formed at the end of the collapse in the framework of classical general relativity, must be removed by quantum effects occurring at the Planck scale. In the present paper, we have shown that a non-minimal coupling of classical gravity with fermions may alter the usual collapse scenario long before quantum gravity effects become important. The qualitative result of the introduction of this non-minimal coupling (and topological term for gravity) is equivalent to the one recently found in a large class of gravity theories in Ref.~\cite{p}. However, at the quantitative level there are substantial differences: here the threshold scale below which black holes do not form might be considerably higher than the Planck mass, and possibly observable. Astrophysical observations might be used to constraint possible allowed values of $\alpha$ and $\gamma$ and thus probe the validity of such a theoretical framework.


{\it Acknowledgments ---}
This work was supported by the NSFC grant No.~11305038,
the Shanghai Municipal Education Commission grant for Innovative 
Programs No.~14ZZ001, the Thousand Young Talents Program, 
and Fudan University.


\end{document}